\documentclass{webofc}

\usepackage[varg]{txfonts}   
\usepackage{hyperref}
\begin{document}
\title{Multi-threaded Output in CMS using ROOT}

\author{\firstname{Daniel} \lastname{Riley}\inst{1}\fnsep\thanks{\email{Daniel.Riley@cornell.edu}} \and
        \firstname{Christopher} \lastname{Jones}\inst{2}
}

\institute{Cornell University
\and
           Fermi National Accelerator Laboratory
          }

\abstract{%
CMS has worked aggressively to make use of multi-core architectures, routinely running 4- to 8-core production jobs in 2017. The primary impediment to efficiently scaling beyond 8 cores has been our ROOT-based output module, which has been necessarily single threaded. In this paper we explore the changes made to the CMS framework and our ROOT output module to overcome the previous scaling limits, using two new ROOT features: the \texttt{TBufferMerger} asynchronous file merger, and Implicit Multi-Threading. We examine the architecture of the new parallel output module, the specific accommodations and modifications that were made to ensure compatibility with the CMS framework scheduler, and the performance characteristics of the new output module.
}
\maketitle
\section{Introduction}
\label{sec:intro}

Several trends have combined to make multi-threading advantageous for processing LHC data.  Increasing LHC event size and complexity have raised the per-event memory and processing requirements.  At the same time the historical increase in CPU clock rates has flattened out, while systems with dozens of processor cores have proliferated.  To meet the experiment's processing goals for memory footprint and event throughput, the design team for the CMS framework has aggressively pursued multi-threading in order to share memory for non-event related data and, more recently, to allow parallel processing within an event~\cite{threadedframework}.

Currently the primary bottleneck limiting scaling of CMS production jobs beyond $O(10)$ cores and threads is writing the data to ROOT format output files, as only one thread at a time can write to a ROOT~\cite{root} format file.  For most CMS output formats, the limiting factor is not the data rate but rather the processing time due to the use of relatively expensive data compression.  Recently the ROOT team has made progress in enhancing the level of parallelism possible in the ROOT I/O subsystem~\cite{rootioparallelism}, and CMS has been exploring the use of these new features in the CMS output module.  In this paper we will discuss the results of these explorations.

\section{ROOT and CMS Framework Multi-Threading}\label{sec:threading}

\subsection{ROOT File Output}\label{sec:rootio}

ROOT serializes C++ objects into a columnar storage format with data compression.  Each column, which may represent an entire C++ object or a sub-field of an object, has its own memory buffer.  All the buffers associated with a file are compressed and written to disk periodically, at a frequency that can be automatically tuned by the ROOT I/O system or set manually.  While ROOT was initially made thread-safe via a ``big lock'' scheme, it is currently in the process of implementing more fine-grained locking and automatic parallelisation.

\subsection{The CMS Data Processing Framework}

The CMS framework processes event data through a sequence of algorithms.  Algorithms are scheduled  ``on demand'': producers of data register what data they consume and produce, and analysis algorithms similarly register the data they consume.  The framework arranges to read data from data sources as needed, schedule producers and analysis Algorithms so that data dependencies are satisfied, and write each event to the output modules when processing on that event is complete.  Conceptually events are processed in ``streams'' that correspond roughly to process threads (threads that are otherwise stalled are allowed to work on other streams when work is available, with the data dependency information used to determine what Algorithms can be run in parallel within an event).

\begin{figure}[htb]
\centering
\includegraphics[width=\textwidth]{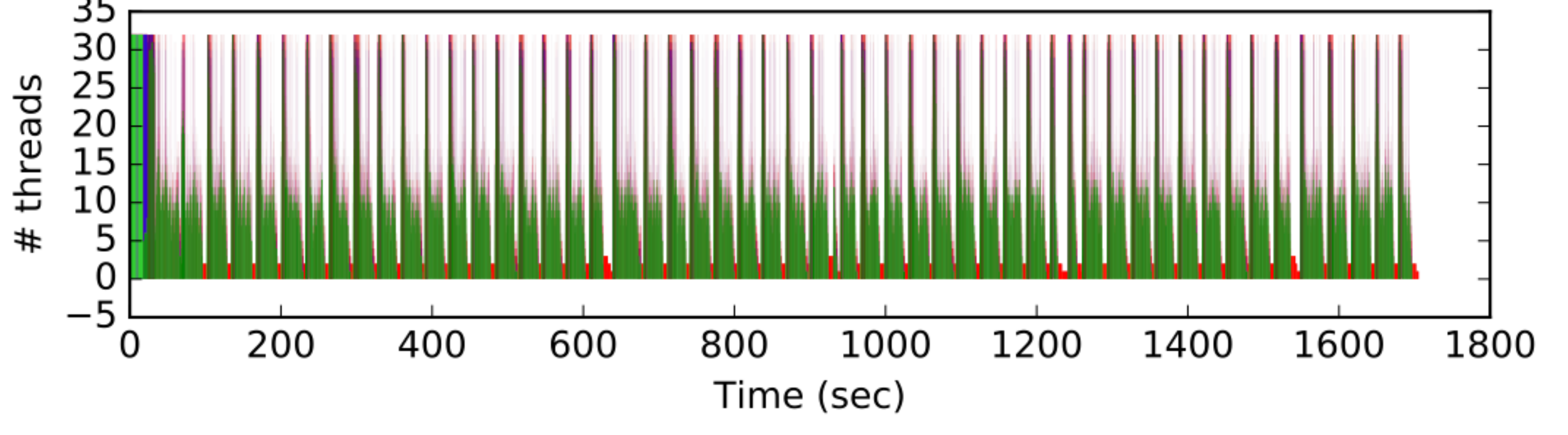}
\caption{Example stall graph for a CMS reconstruction job with 32 threads.  Perfect efficiency is achieved when the number of modules running concurrently equals the number of threads.  In the figure, dark green shows concurrent events, gaps filled with red represent inefficiencies where the output module is stalled.}
\label{fig:recostallgraph}
\end{figure}

Most of the Algorithms used in CMS production jobs have been made thread safe and capable of concurrent processing, typically by allowing independent instances of the Algorithm for each stream.  The most notable exception is the module that writes events to ROOT-format output files due to the limitations discussed in the previous sections.  In order to assess the levels of concurrency achievable with the CMS framework, we have developed our own internal stall monitoring system that samples and records the level of concurrency achieved in any CMS processing job.  An example ``stall graph'' for a typical reconstruction task configured to use 32 threads (details of the job are documented in Section~\ref{sec:performance}) is shown in Figure~\ref{fig:recostallgraph}.  Gaps in the stall graph represent inefficiencies where threads are idle while waiting on the completion of a non-concurrent module.  Detailed monitoring shows that most of the gaps are primarily due to the ROOT output module, with the periodicity of the gaps showing the frequency with which buffers are compressed and written to disk.  Profiling of the output module reveals that the primary bottleneck, particularly for the analysis data formats, is the processor time compressing the buffers, not the actual writes to storage.

\section{CMS use of ROOT I/O Parallelism}

The ROOT developers have recently implemented several improvements to the parallelism possible in the ROOT IO system~\cite{rootioparallelism}.  CMS has investigated how two of these developments, \textit{Implicit Multi-Threading} (IMT) and the \texttt{TBufferMerger} class, can be used to improved the scalability of CMS jobs to more than $O(10)$ threads.

\subsection{Implicit Multi-Threading}\label{sec:imt}

As explained in Section~\ref{sec:rootio}, data elements (entire classes or sub-fields of a class) are serialized into independent buffers that are periodically compressed and written to storage.  Implicit Multi-Threading~\cite{rootioparallelism} parallelizes the loops over the buffers in the serialization and compression steps.

In principle, using IMT is as simple as setting an option enabling it.  In practice, we found subtle interactions due to the threading system used by both CMS and ROOT, Intel Threading Building Blocks (TBB)~\cite{tbb}.  TBB uses a work-stealing system where idle threads can steal work from the task queues of busy threads.  When IMT is enabled, the output module thread creates a set of tasks, one for each buffer to be compressed (for instance), on its own work queue.  Idle threads may steal work from the output module's thread; when the output module thread exhausts its task queue, it waits for any stolen tasks to complete before proceeding. While the output module thread is waiting, it may steal work from another thread's queue, possibly stealing a time-consuming CMS task leading to the output module thread blocking until the expensive CMS task completes.  To avoid the inefficiencies this could create, we found it necessary to use TBB's task isolation feature to prevent the output module thread stealing work created by other threads.

We found IMT to be most effective where there are many data columns of similar size and expensive compression is applied, as this combination gives IMT the most work to do and a balanced set of tasks.  IMT was less effective when the data to be written created few data columns or the column sizes were severely unbalanced.

\subsection{TBufferMerger}

The \texttt{TBufferMerger}~\cite{rootioparallelism} approach avoids the single-thread file output limitation by creating many data buffers, based on the ROOT \texttt{TMemFile} memory file object, that are merged to the final output file.  Serialization and compression are performed on the data buffers and hence can be parallelized.  Conceptually, this allows parallelization at a coarser granularity than IMT.

The simplest use of \texttt{TBufferMerger} would be to create a buffer for each stream/thread.  CMS decided on a somewhat more complicated scheme for several reasons:
\begin{itemize}
\item Each buffer takes a significant amount of memory, so creating a buffer for every thread conflicts with our goal of maximizing memory sharing to minimize memory requirements.
\item Using many buffers increases the ``tail effects'' at the end of the job where most buffers are only partially filled, which degrades compression and file layout optimization.
\item Filling multiple buffers at approximately the same rate can lead to synchronization effects, where multiple buffers fill up at the same time, leading to contention and stalls writing to the queue for the merge step.
\end{itemize}
To mitigate these problems, CMS implemented a new module type that allows a configurable limit on the module's concurrency in the framework scheduler that is used for the CMS parallel output module.  This parallel output module uses \texttt{TBufferMerger}, creating as many output memory buffers as the concurrency limit.  Internally, the buffers are stored in a TBB concurrent priority queue, with the priority set such that when an event is written it goes to the available buffer with the most events already stored.  Filling the fullest available buffer minimizes synchronization and tail effects.

With the version of \texttt{TBufferMerger} used for these tests, the merge operation occurs on the thread of the parallel output module that triggers the merge.  We found that merge operation could take noticeable processor time, so it was necessary to keep the merge operation within the scope of the framework's task management to avoid using more than the allotted computational resources.

\section{Performance Tests}\label{sec:performance}

\subsection{Test Configurations}

For the performance tests we used simulated $t\bar{t}$ data with LHC run2 conditions.  The full reconstruction step was run, with two output scenarios:
\begin{enumerate}
	\item Full reconstruction output (RECO), analysis oriented output (AOD), and MINIAOD
	\item AOD and MINIAOD
\end{enumerate}
Two platforms were tested:
\begin{enumerate}
	\item 32 core Skylake-SP Gold 6130 CPU @ 2.10 GHz
	\item 64 core Xeon Phi Knights Landing KNL 7210 @ 1.30 GHz
\end{enumerate}
For each combination of output scenario and platform, we tested these processing configurations:
\begin{enumerate}
	\item Standard single-threaded output module with IMT disabled
	\item Standard single-threaded output module with IMT enabled
	\item Parallel output module using \texttt{TBufferMerger} with IMT, with RECO output module concurrency 6, AOD concurrency 6, and MINIAOD concurrency 3
	\item A dummy output module that does no event serialization or compression to test the framework's scalability with a perfectly scalable output module
\end{enumerate}

\subsection{Results}

\begin{figure}[htb]
\centering
\includegraphics[width=\textwidth]{recostallgraph.pdf}
\includegraphics[width=\textwidth]{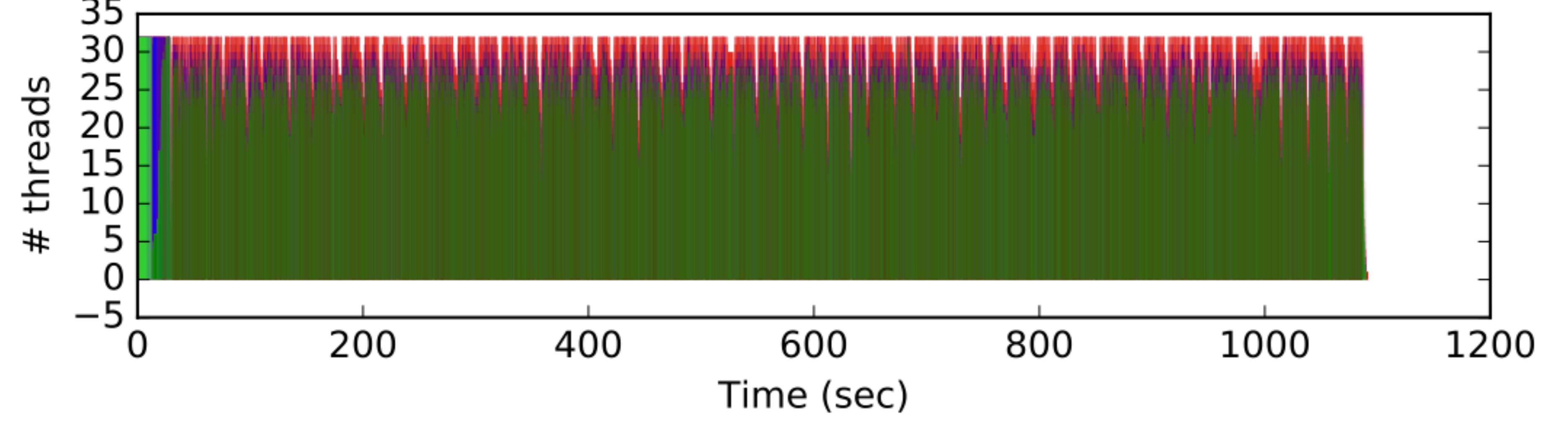}
\caption{Stall graphs comparing the standard output module with IMT disabled and the parallel output module with IMT enabled showing a substantial reduction in stalls and total processing time (note the change of horizontal scale).  Tests were writing RECO/AOD/MINIAOD (output scenario 1) on the Skylake-SP system (platform 1).  The stall graph represents only CMS module concurrency, so it does not account for IMT ``filling in the gaps'' by stealing work.  As a result, the stall graph underestimates the actual CPU efficiency.}
\label{fig:recostallgraphcompare}
\end{figure}

Figure~\ref{fig:recostallgraphcompare} compares the stall graphs for the standard output module with IMT disabled and the parallel output module with IMT enabled.  The stall graphs show a one-third reduction in total processing time, from $\sim 1700$ to $\sim 1100$ seconds, and a significant reduction in stalls.

\begin{figure}[htb]
	\includegraphics[width=0.45\textwidth]{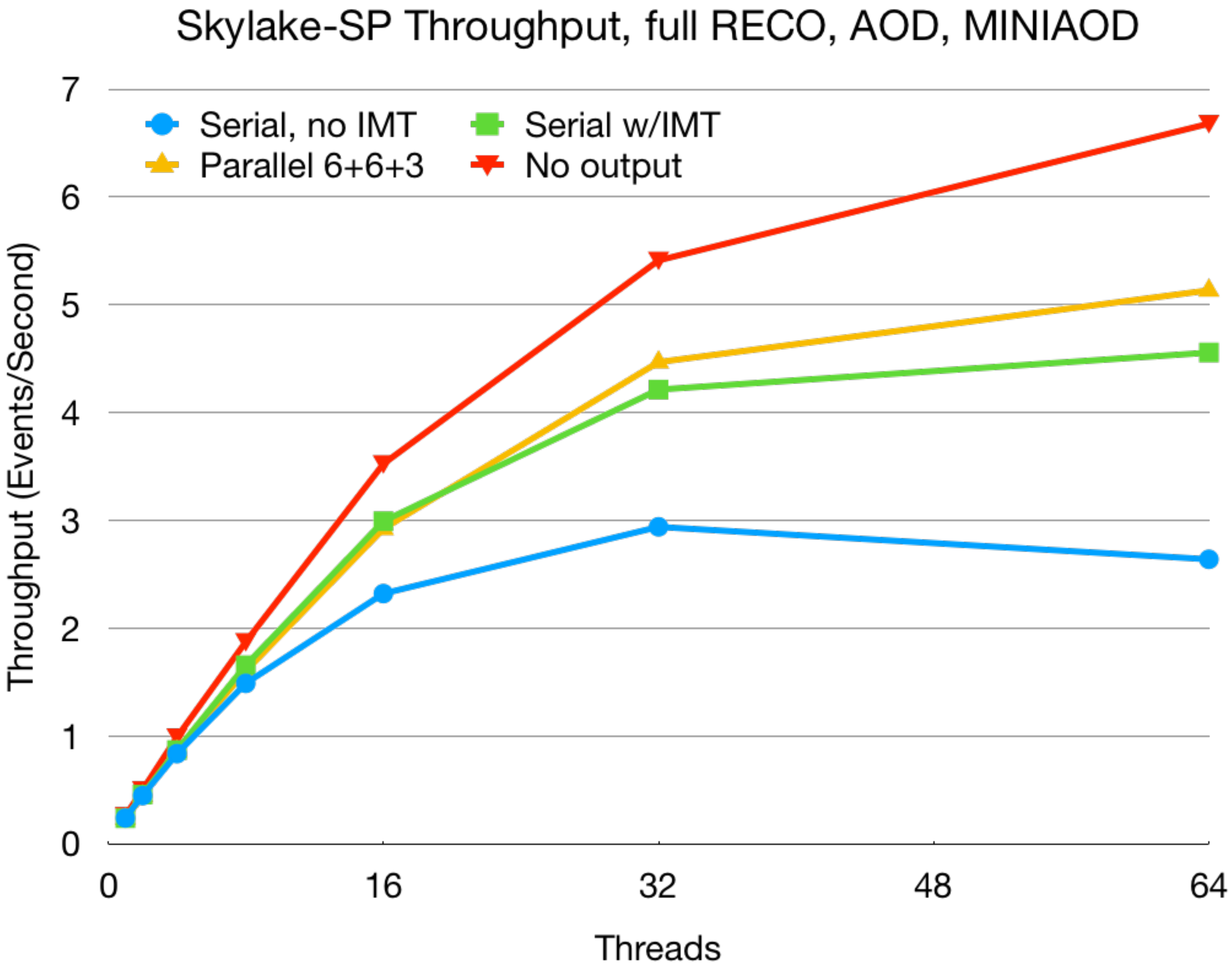}
	\hfill
  \includegraphics[width=0.45\textwidth]{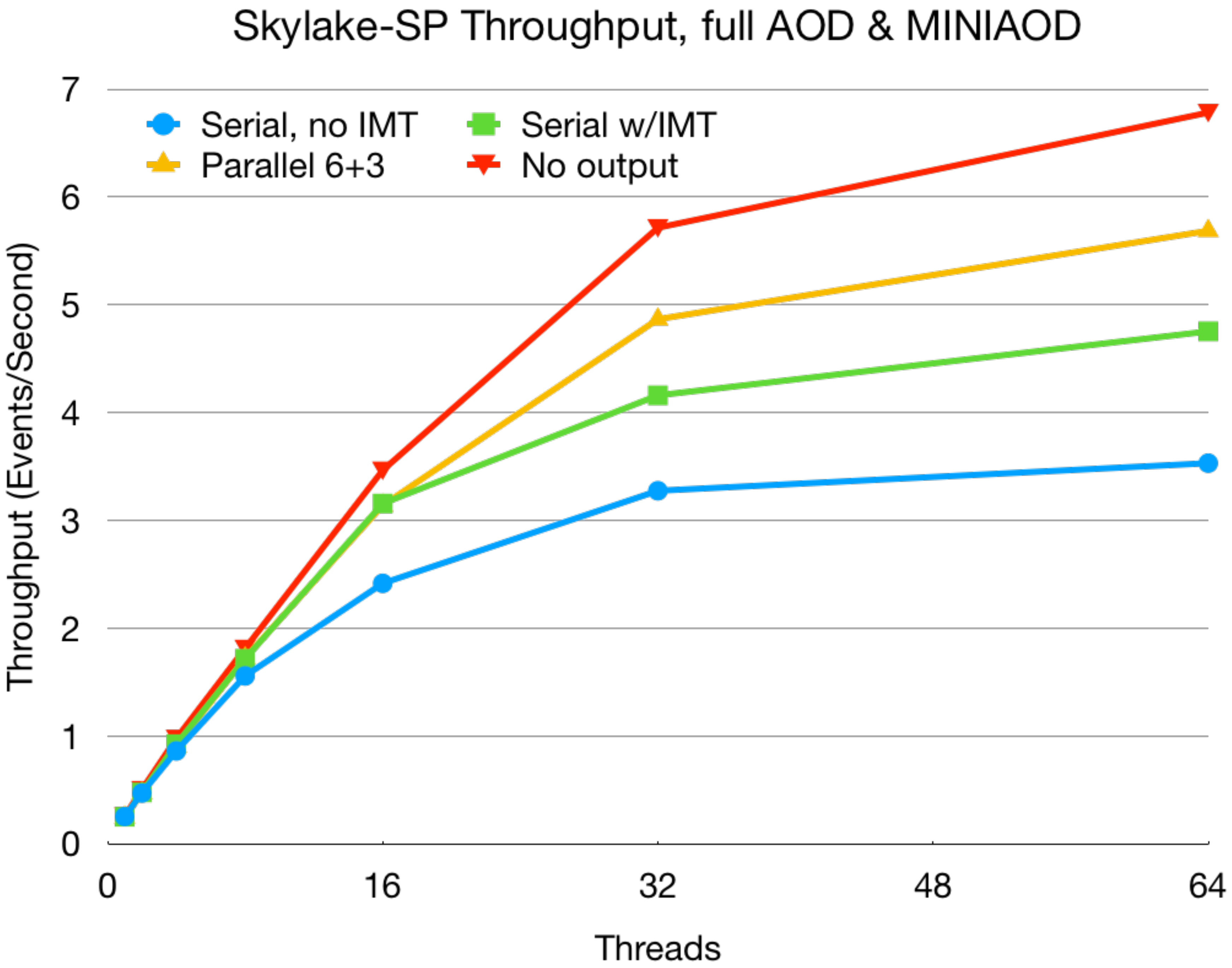}
 	\caption{Scaling behavior of the different processing configurations on Skylake-SP, writing RECO/AOD/MINIAOD (left) and AOD/MINIAOD (right).}
 	\label{fig:skylakecomp}
\end{figure}

Figure~\ref{fig:skylakecomp} compares the scalability performance for the different processing configuration on the Skylake-SP platform.  For both configurations enabling IMT gives a substantial speedup, with the parallel output module only offering further improvement at high concurrency levels.  IMT does particularly well for the RECO/AOD/MINIAOD scenario, where the large volume and number of products in the RECO output gives IMT more opportunities for parallelization.  Profiling provides some evidence that failure to reach the framework's overall scaling limits are at least partially due to lock contention inside ROOT that the developers are currently addressing.

\begin{figure}[htb]
	\centering
  \includegraphics[width=0.45\textwidth]{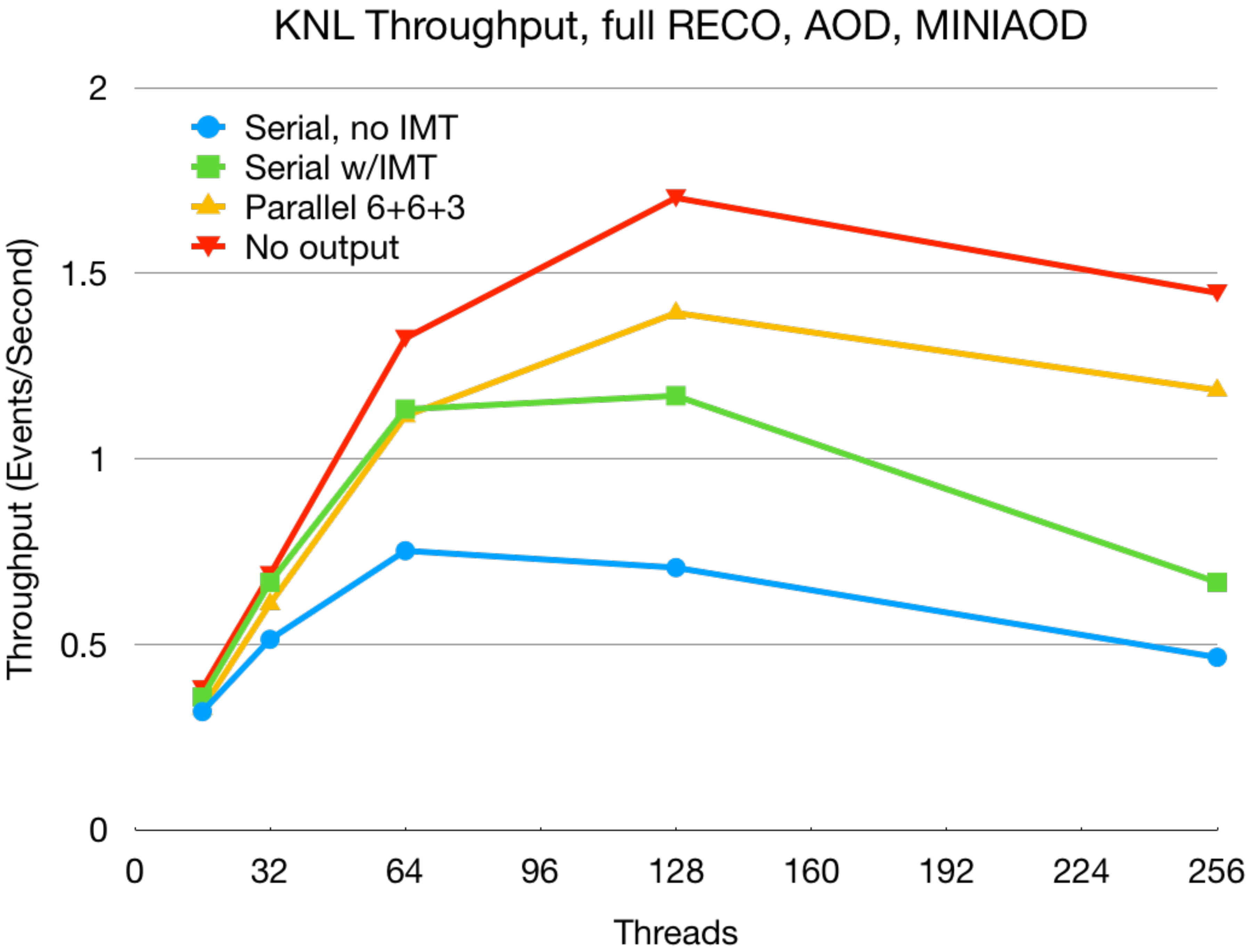}
 	\caption{Scaling behavior of the different processing configurations on Knights Landing, writing RECO/AOD/MINIAOD.}
 	\label{fig:knlcomp}
\end{figure}

Figure~\ref{fig:knlcomp} shows the RECO/AOD/MINIAOD scaling on the KNL platform.  As in the Skylake-SP tests, the parallel output module follows the shape of the framework scaling limits past the point where IMT alone runs out of work to parallelize.

\section{Conclusions}

The parallel I/O features in recent versions of ROOT have the potential to significantly improve the scalability of the CMS output module, which in turn improves our overall job scalability.  After addressing the threading subtleties discussed in Section~\ref{sec:imt}, CMS has enabled IMT by default, which significantly improves our scalability for concurrency levels in the range from 10-20 threads.  We are currently working on completing some loose ends in the implementation of the parallel output module, which promises to improve our job scalability further at high concurrency levels.

\end{document}